\documentclass[osajnl,twocolumn,showpacs,superscriptaddress,10pt]{revtex4-1} 

\usepackage{amsmath,amssymb,graphicx}

\begin{document}

\title{Collimated Blue and Infrared Beams Generated by Two-Photon Excitation in Rb Vapor}


\author{J. F. Sell}\email{Corresponding author: jerry.sell@usafa.edu}
\affiliation{Laser and Optics Research Center, Department of Physics, U.S. Air Force Academy, Colorado 80840, USA}

\author{M. A. Gearba}
\affiliation{Laser and Optics Research Center, Department of Physics, U.S. Air Force Academy, Colorado 80840, USA}
\affiliation{Department of Physics and Astronomy, University of Southern Mississippi, Hattiesburg, Mississippi 39406, USA}

\author{B. D. DePaola}
\affiliation{Department of Physics, Kansas State University, Manhattan, Kansas 66506, USA}

\author{R. J. Knize}
\affiliation{Laser and Optics Research Center, Department of Physics, U.S. Air Force Academy, Colorado 80840, USA}

\begin{abstract}{\vspace{-5ex}}
Utilizing two-photon excitation in hot Rb vapor we demonstrate the generation of collimated optical fields at 420 nm and 1324 nm.  Input laser beams at 780 nm and 776 nm enter a heated Rb vapor cell collinear and circularly polarized, driving Rb atoms to the $5D_{5/2}$ state.  Under phase-matching conditions coherence among the $5S_{1/2} \rightarrow 5P_{3/2} \rightarrow 5D_{5/2} \rightarrow 6P_{3/2}$ transitions produces a blue (420 nm) beam by four-wave mixing.  We also observe a forward and backward propagating IR (1324 nm) beam, due to cascading decays through the $6S_{1/2} \rightarrow 5P_{1/2}$ states.  Power saturation of the generated beams is investigated by scaling the input powers to greater than 200 mW, resulting in a coherent blue beam of 9.1 mW power, almost an order of magnitude larger than previously achieved.  We measure the dependences of both beams in relation to the Rb density, the frequency detuning between Rb ground state hyperfine levels, and the input laser intensities.
\end{abstract}


\maketitle 


A wide range of phenomena can be created by exploiting nonlinear optical processes in a dense atomic vapor.  Large enhancements of these processes are possible through the generation of quantum coherences among atomic states and include effects such as electromagnetically induced transparency, fast and slow light propagation, four-wave mixing, and lasing without inversion.  Four-wave mixing (FWM) in particular has been shown to produce both efficient frequency up-conversion \cite{Akulshin2009,Meijer2006,Vernier2010} and down-conversion \cite{Becerra2008} using low power continuous wave (cw) lasers.  The newly created optical fields are narrowband tunable coherent light sources \cite{Akulshin1-2012}, with wavelengths from the IR to approaching the UV depending upon the atomic states involved.

\begin{figure}
\includegraphics{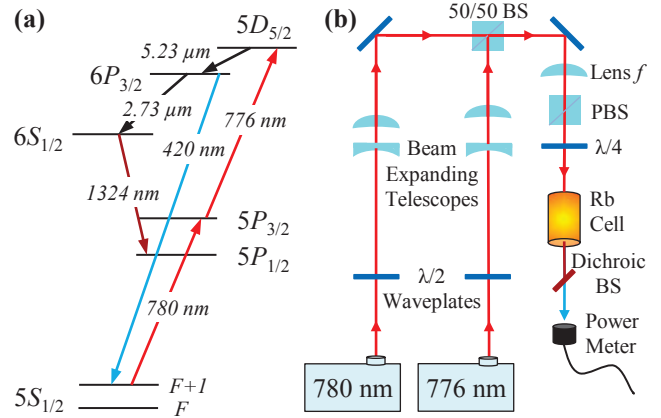}
\caption{(Color online) (a) Relevant Rb energy levels.  Excitation lasers (cw) are present at 780 nm and 776 nm while collimated output beams are observed at 420 nm and 1324 nm. (b) Schematic of the experimental setup.}
\end{figure}

Frequency up-conversion by FWM has most often been studied in Rb vapor, first demonstrated using low power cw lasers by Zibrov \textit{et al.} in 2002 \cite{Zibrov2002} where 15 $\mu$W of coherent radiation at 420 nm was achieved.  The method relies upon input lasers at 780 nm and 776 nm [see Fig. 1(a)] driving Rb atoms from the $5S_{1/2}$ ground state to the  $5D_{5/2}$ state by two photon excitation, with the $5P_{3/2}$ level as an intermediate state.  A third optical field between the $5D_{5/2} \rightarrow 6P_{3/2}$ levels at 5.23 $\mu$m is initiated through spontaneous emission.  Strong atomic coherences are thus formed in a diamond-type energy level structure, creating coherent blue light at 420 nm ($6P_{3/2} \rightarrow 5S_{1/2}$) by FWM.  Recent experiments achieved first 40 $\mu$W of 420 nm light through the additional coupling of both $5S_{1/2}$ hyperfine ground state levels \cite{Akulshin2009}, and subsequently 1.1 mW by further optimization of input laser polarizations and frequencies \cite{Vernier2010}.  The generated blue beam exhibits a high degree of spatial coherence \cite{Meijer2006}, with a spectral linewidth typically limited by the linewidths of the applied laser fields \cite{Akulshin2013}.  The absolute frequency of the blue light has been found to be centered on the $6P_{3/2} \rightarrow 5S_{1/2}$ transition, with tunability of $\geq$ 100 MHz possible by adjustment of the input laser frequencies \cite{Akulshin1-2012}.  Incorporating an additional laser at 795 nm has been shown to both enhance and suppress the FWM process through control of optical pumping \cite{Akulshin2-2012}.  A single frequency laser at 778 nm has also been shown to produce blue (420 nm) beams, using both cw \cite{Brekke2013} and pulsed excitation \cite{Sulham2010}.  Utilizing the corresponding two-step excitation scheme in Cs, 4 $\mu$W of coherent blue light was generated at 455 nm \cite{Schultz2009}, illustrating the application of this method to less ideal cases as only 0.4\% of Cs atoms cascade from $6D_{5/2} \rightarrow 7P_{3/2}$ compared to 35\% of Rb $5D_{5/2}$ atoms which decay through the $6P_{3/2}$ state \cite{Heavens1961}.

In this Letter we analyze the collimated beams at 420 nm and 1324 nm which are produced during two-photon excitation to the Rb $5D_{5/2}$ state.  Substantially greater blue beams than previously observed are achieved by scaling the input beams to much higher intensities while adjusting their frequency detunings along with the Rb density in order to obtain optimal phase-matching and input beam absorption for blue beam generation and transmission.  The coherent and collimated blue light can be used as a basis for applications such as sensitive atom detection \cite{Akulshin2011}, quantum information processing \cite{Walker2012}, and underwater communication \cite{Sulham2010}.  The generated 1324 nm beam has seldomly been studied; however, we find it can be significant reaching levels of 2.5 mW or more.  The forward and backward propagating 1324 nm beam can occur independently or simultaneously with the generated blue beam, which can help to characterize the coherence and population dynamics occurring during the FWM process.

The experimental setup is illustrated schematically in Fig. 1(b).  A master oscillator and tapered amplifier laser system drives the $D_{2}$ transition in Rb at 780 nm, while a titanium:sapphire ring laser drives the $5P_{3/2} \rightarrow 5D_{5/2}$ transition at 776 nm.  The two laser beams are combined in a 50/50 non-polarizing beamsplitter and pass through a $\lambda$/4 waveplate in order to achieve co-propagating and circularly polarized input optical fields, whose relevance in coherent blue light generation has been identified in previous experiments \cite{Akulshin2009,Vernier2010}.  An $f = 250$ mm lens is used to focus the laser beams to a waist radius of $\simeq$ 100 $\mu$m in the vapor cell, resulting in intensities of up to $10^{3}$ W/cm$^{2}$ (350 mW power) in each laser beam.  This choice was made to allow a depth of focus approximately the length of the vapor cell (5 cm).  The Rb vapor cell contains a natural abundance of Rb and is operated from $105 - 135 ^{\circ}$C ($8 \times 10^{12} - 5 \times 10^{13}$ cm$^{-3}$).  Beam expanding telescopes are also incorporated into each laser beam path to optimally overlap the beams in the Rb vapor cell.  We measure the power of the generated beams after they pass through a dichroic beamsplitter which reflects $\geq99.8\%$ of the light from $776 - 780$ nm, while transmitting $\geq75\%$ at 420 nm and 1324 nm.  Possible beam emissions at 2.7 $\mu$m and 5.2 $\mu$m are not observed as they do not transmit through the windows of the vapor cell.

\begin{figure}
\includegraphics{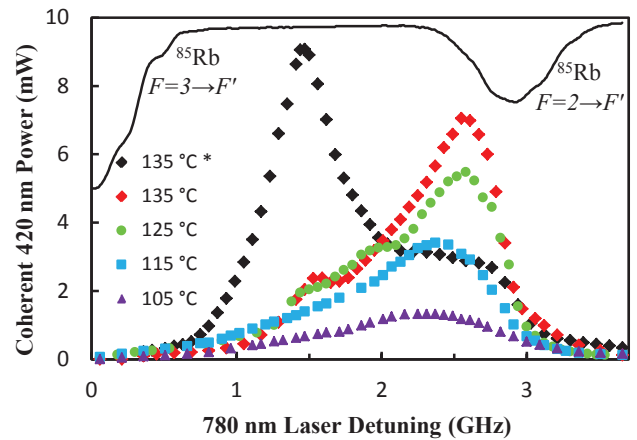}
\caption{(Color online) The generated blue beam power as a function of the 780 nm laser frequency, referenced to $^{85}$Rb ground hyperfine state absorption peaks (shown in the upper trace). Vapor cell temperatures are given, with * denoting a shift in the laser beams foci to the entrance of the vapor cell.}
\end{figure}

Figure 2 illustrates the coherent 420 nm power generated as a function of the 780 nm laser detuning from the $^{85}$Rb $5S_{1/2} (F=3) \rightarrow 5P_{3/2} (F')$ absorption resonance.  A Burleigh WA-1500 wavemeter (resolution of 0.001 cm$^{-1}$ $\simeq$ 30 MHz) is used to record the 780 nm laser frequency, which is calibrated by scanning across Rb absorption features in a room temperature vapor cell (upper trace in Fig. 2).  The 776 nm frequency is also monitored with this wavemeter, where the 776 nm laser detuning ($\delta _{776}$) is correspondingly adjusted relative to the 780 nm laser detuning ($\delta _{780}$) to remain on the $5D_{5/2}$ resonance ($\delta _{780} = - \delta _{776}$).  A linear rise in the blue beam power is observed with increasing Rb vapor cell temperatures, assuming increasing input powers are available to saturate the FWM process.  We examine these saturation levels later in Fig. 3.  A threshold Rb vapor cell temperature of 95$^{\circ}$C is extrapolated for blue beam generation under our experimental conditions, while 135$^{\circ}$C is the highest temperature in Fig. 2, as the blue power begins to decrease above this value.  Using 135$^{\circ}$C as an approximately optimum vapor cell temperature with respect to our maximum input beam powers, we proceeded to empirically test different focusing lens combinations in order to achieve the greatest 420 nm power.  This occurred when the foci of the input laser beams were shifted from the center to the entrance of the vapor cell, denoted by the * data set in Fig. 2.  A 420 nm beam of 9.1 mW power was obtained, with input beam diameters of $\simeq$ 85 $\mu$m (FWHM) at the entrance of the vapor cell and $\simeq$ 310 $\mu$m at the exit, using input laser powers of $P_{780} = 390$ mW and $P_{776} = 205$ mW.

These results can be qualitatively understood by considering the phase matching conditions, $\vec{k}_{780} + \vec{k}_{776} = \vec{k}_{420} + \vec{k}_{5230}$ for this FWM process, with the wave vector magnitudes $k_{\omega} = \omega /c \cdot n_{\omega}$, where $n_{\omega}$ is the refractive index at the respective frequency.  The 780 nm and 420 nm optical fields encounter the largest refractive index variation as they are readily absorbed by ground state Rb atoms.  However, phase-matching will be most favorable for minimal changes in $n_{780}$ and $n_{420}$, which Meijer \textit{et al.} \cite{Meijer2006} calculated for low laser powers to occur at 780 nm detunings from $1-3$ GHz.  Our data agree well with this result at $\geq 10$ times the input laser intensities.  Small changes in the optimum 780 nm detunings are observed as the Rb density and input laser intensities are increased.  This can be attributed to slight changes in $n_{780}$ and $n_{420}$ which serve to optimize phase-matching as the laser detuning moves along the wing of the Doppler broadened $^{85}$Rb $5S_{1/2} (F = 2) \rightarrow 5P_{3/2} (F')$ resonance.

Shifting the input beam focus to the entrance of the vapor cell not only resulted in a significant increase in 420 nm output power but also in a large shift in the 780 nm detuning.  We note this detuning agrees with results at lower powers which determined ideal 780 nm detunings almost directly between the $^{85}$Rb ground hyperfine states \cite{Akulshin2009,Vernier2010}.  Propagation effects such as this have not been well studied; however, similar experiments in sodium highlighted the competition between FWM and amplified spontaneous emission (ASE) in two-photon excitation \cite{Boyd1987}.  It was suggested that when the beam focus is moved toward the entrance of the vapor cell, FWM is initiated more strongly, leading to suppression of upper level populations and associated ASE.  Another noticeable propagation effect is an increase in the generated 420 nm beam power by $5 - 10\%$ when blocking the edges of the input laser beams with an aperture.  While we would not expect the far wings of the spatially Gaussian beams to contribute to FWM it is interesting that they slightly decrease the process, possibly also due to increased competition from ASE.

Figure 3 illustrates the generated blue beam power as a function of each input beam power for vapor cell temperatures from $105 - 135 ^{\circ}$C.  As the 780 nm input power is increased [Fig. 3(a)], we observe both a threshold-like behavior along with saturation of the blue beam power.  These characteristics can be described by considering the optical depth (OD) for the 780 nm beam.  The optical depth, OD $= N \sigma l$, where $N$ is the Rb density, $\sigma$ is the absorption cross section, and $l$ is the length of the vapor cell.  In calculating $\sigma$ we take into account the laser intensities and frequency detunings using
\begin{equation}
	\sigma = \frac{\sigma_{0}}{1+4(\Delta/\Gamma)^{2}+(I/I_{sat})},
\end{equation}
where $\sigma_{0}$ is the on-resonance cross section, $\Delta$ is the detuning from resonance, $\Gamma$ is the spontaneous decay rate, $I$ is the 780 nm laser intensity, and $I_{sat}$ is the saturation intensity for the $5P_{3/2}$ transition.  Parameters such as $\sigma_{0}$ and $I_{sat}$ along with Rb vapor pressures are taken from the data compiled by Steck \cite{Steck2013}.  Using the results shown in Fig. 3(a), we extract the 780 nm laser intensities at the onset of blue light generation and when the blue power saturates.  We find, regardless of temperature, the blue power saturates at OD $\simeq$ 1.  At higher levels of absorption (corresponding to smaller 780 nm input powers) the blue power decreases until OD $\simeq 5 - 6$ where a blue beam is no longer observed.  In this circumstance, the 780 nm beam is quickly absorbed in the initial length of the vapor cell, at which point the FWM process ceases and the generated 420 nm beam is also absorbed before it can exit the vapor cell.  Figure 3(b) does not show a threshold type of behavior as the 780 nm beam is already at saturation intensities, allowing the FWM process to propagate throughout the Rb vapor cell even at low 776 nm intensities as the 776 nm beam is weakly absorbed in the Rb vapor.  As the above results are consistent across a range of temperatures they may be used to estimate an optimum Rb density depending upon the laser intensities available.

\begin{figure}
\centerline{\includegraphics{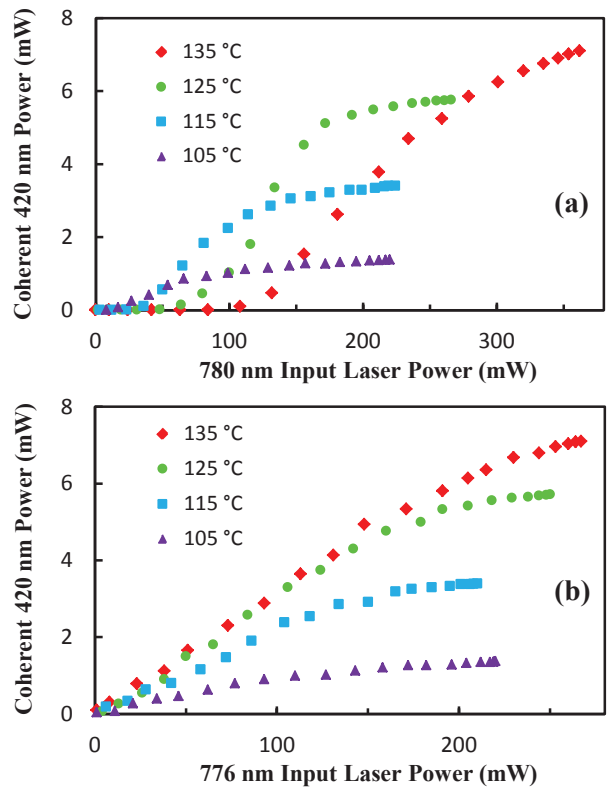}}
\caption{(Color online) (a) Output blue power measured while varying 780 nm input power and holding the 776 nm input power at its maximum value.  (b) Blue power measured while varying the 776 nm input power and holding the 780 nm input power at its maximum value.}
\end{figure}

\begin{figure*}
\includegraphics{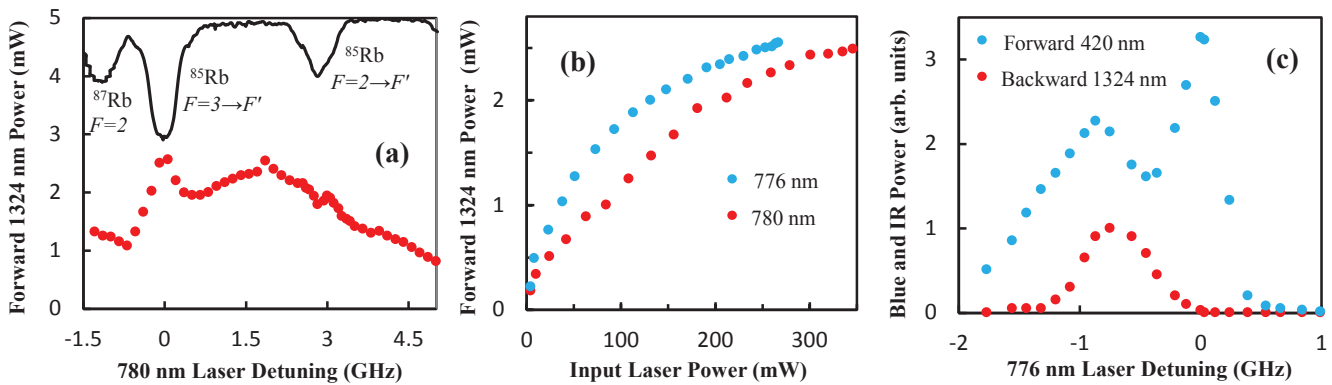}
\caption{(Color online) (a) Forward IR beam power as the input 780 nm laser is scanned in frequency.  The upper curve references our laser detunings to Rb absorption resonances. (b) Forward IR power dependence on input laser powers (as one input beam power is adjusted the other is held constant at its maximum power). (c) The forward blue and backward IR beams measured simultaneously as the 776 nm frequency is varied.}
\end{figure*}

During analysis of the generated blue beam we also observed a collimated IR beam, measured to be at 1324 nm using an optical spectrum analyzer.  Examining this beam with an InGaAs scanning slit beam profiler showed a Gaussian spatial profile.  Previous studies by Zibrov \textit{et al.} \cite{Zibrov2001} observed a 1.36 $\mu$m beam which they accounted for by FWM through the $6S_{1/2} \rightarrow 5P_{3/2}$ states.  A recent paper confirmed such a forward beam, while also inferring the presence of a 1.32 $\mu$m beam \cite{Akulshin2013-2}.  In our case the only wavelength detected is 1324 nm, corresponding to the $6S_{1/2} \rightarrow 5P_{1/2}$ transition.  Inserting a dichroic beamsplitter before the vapor cell which transmits $776-780$ nm but reflects 1.3 $\mu$m revealed this beam is both forward and backward propagating.  This suggests a process involving population transfer such as amplified spontaneous emission and rules out FWM since phase-matching dictates a forward-only beam.

Figure 4 illustrates the dependence of the 1324 nm beam depending on the input lasers' frequencies and powers.  As phase-matching is not required to produce the 1324 nm beam, Fig. 4(a) shows a much greater range of 780 nm detunings where this beam is generated compared to the blue beam in Fig. 2.  Figure 4(b) demonstrates the power in the forward component of the 1324 nm beam, which is close to reaching saturation only at the highest input powers.  By scanning the 776 nm laser frequency while simultaneously monitoring both the forward only 420 nm beam and the backward propagating beam at 1324 nm we obtain the data shown in Fig. 4(c).  The 1324 nm beam is observed over a large detuning range where the blue beam occurs, suggesting cascading population inversions on the $5D_{5/2}$, $6P_{3/2}$, and $6S_{1/2}$ states in addition to the FWM process, in agreement with recent results observing the 5.2 $\mu$m beam emission \cite{Akulshin2013-2}.  By filtering between the blue and IR beams, we measure a forward beam at 1324 nm of 1.2 mW when producing 9.0 mW at 420 nm.  This does not correspond to the maximum or minimum 1324 nm power, demonstrating the interplay between population dynamics and FWM processes at work.  A complete model to describe this system will need to include the coherence and population among the six atomic states and their sublevels.

While we have examined the generated blue and IR beams in terms of the input laser intensities, Rb density, and laser frequency detunings, a theory is yet to be developed which provides an upper limit on the efficiency of this process.  We estimate a 40\% increase in the blue beam power could be achieved simply by using anti-reflection coated vapor cells and optimized dichroic optics when splitting the output beams.  Further advances have the potential to not only increase the efficiency of frequency up and down conversion in Rb but to generalize the characteristics of this process to other states and atoms which have wavelengths of interest.

The authors gratefully acknowledge support of this research by the Air Force Office of Scientific Research and the National Science Foundation (grant 1206128).

\end{document}